\documentclass[english]{aastex}
\usepackage[T1]{fontenc}
\usepackage[latin1]{inputenc}
\usepackage{graphicx}

\makeatletter

\newcommand{\noun}[1]{\textsc{#1}}
\providecommand{\tabularnewline}{\\}

\newcommand{\eg}{{\it e.g.\ }}
\newcommand{\ie}{{\it i.e.\ }}

\newcommand{\ip}{i^{\prime}}
\newcommand{\rp}{r^{\prime}}
\newcommand{\gp}{g^{\prime}}
\newcommand\las{\mathrel{\hbox{\rlap{\hbox{\lower3pt\hbox{$\sim$}}}\hbox{\raise2pt\hbox{$<$}}}}}
\newcommand\gas{\mathrel{\hbox{\rlap{\hbox{\lower3pt\hbox{$\sim$}}}\hbox{\raise2pt\hbox{$>$}}}}}

\usepackage{amssymb}

\usepackage{babel}
\makeatother
\begin{document}

\shorttitle{Main belt asteroids}

\shortauthors{Wiegert et al.}

\title{Evidence for a colour dependence in the size distribution of main belt asteroids}

\author{Paul Wiegert}

\affil{Dept. of Physics and Astronomy, The University of Western Ontario, London, Ontario N6A 3K7 CANADA}
\email {pwiegert@uwo.ca}

\author{David Balam}

\affil{Dept. of Physics and Astronomy, The University of Victoria, Victoria, British Columbia V8W 3P6 CANADA}

\author{Andrea Moss}

\affil{Dept. of Physics and Astronomy, The University of Western Ontario, London, Ontario  N6A 3K7 CANADA}

\author{Christian Veillet}

\affil{Canada-France-Hawaii Telescope Corporation, Kamuela, Hawaii  96743 USA}

\author{Martin Connors}

\affil{Centre for Science, Athabasca University, Athabasca, Alberta T6G 0R9 CANADA}

\author{Ian Shelton}

\affil{Centre for Science, Athabasca University, Athabasca, Alberta, CANADA}

\begin{abstract}
We present the results of a project to detect small ($\sim$1 km) main-belt (MB) asteroids with the 3.6 meter Canada-France-Hawaii Telescope (CFHT). We observed in 2 filters (MegaPrime $\gp$ and $\rp$) in order to compare the results in each band. Owing to the observational cadence we did not observe the same asteroids through each filter and thus do not have true colour information. However strong differences in the size distributions as seen in the two filters point to a colour-dependence at these sizes, perhaps to be expected in this regime where asteroid cohesiveness begins to be dominated by physical strength and composition rather than by gravity. The best fit slopes of the cumulative size distributions (CSDs) in both filters tend towards lower values for smaller asteroids, consistent with the results of previous studies. In addition to this trend, the size distributions seen in the two filters are distinctly different, with steeper slopes in $\rp$ than in $\gp$.  Breaking our sample up according to semimajor axis, the difference between the filters in the inner belt is found to be somewhat less pronounced than in the middle and outer belt, but the CSD of those asteroids seen in the $\rp$ filter is consistently and significantly steeper than in $\gp$ throughout. The CSD slopes also show variations with semimajor axis within a given filter, particularly in $\rp$. We conclude that the size distribution of main belt asteroids is likely to be colour dependent at kilometer sizes and that this dependence may vary across the belt.

\end{abstract}

\keywords{minor planets, asteroids; solar system, general}

\section{Introduction}

Observations of small (of order 1 km diameter) main-belt asteroids
(MBAs) present a considerable challenge. As a result, this faint (typically
$V\geq22$) population of asteroids has not been sampled as well as
it might be. The asteroid size distribution in the main-belt is affected
by a number of factors, the most important of which is thought to
be collisions with other asteroids. It is well known that if the bodies
are uniform in composition and respond to collisions in size-independent
way (\ie have the same strength to mass ratio), the differential
size distribution in steady-state is independent of the details of
the collisions, and is given by a power-law \begin{equation}
dN\propto D^{-p}dD\end{equation}
 where $D$ is the diameter, $dN$ the number of bodies in the size
range $D$ to $D+dD$ and the index $p=3.5$ \citep{doh69}. 

This description is an idealization: in reality, asteroids are affected
by size-dependent phenomena (\eg the Yarkovsky effect, size-dependent
internal strength) and are not in a true steady-state (material leaves
the belt through orbital resonances). It is expected rather that the
main belt will show a depletion of bodies at smaller sizes (which
have less gravitational reinforcement than larger bodies and are hence
more fragile per unit mass, as well as being more quickly removed
by Yarkovsky forces), and that the ideally featureless power-law slope
may display {}``waves'' as a result of these removal processes \citep{davryafar94,durgrejed98,obrgre03}. 

In this paper, the size distribution of the main belt at kilometer
to sub-kilometer sizes is measured in two filters, in order to extend
our knowledge into the regime ($D\las1~$km) where internal strength
plays an increasingly important role in the bodies' response to collisions
\citep{farpaozap82,houhol90,houhol99,benasp99}, and where compositional
differences (possibly indicated by colour differences) become increasingly
important to asteroid cohesiveness and strength.

\section{Observations and data reduction}

All images were taken with the MegaPrime/MegaCam camera on 3.6 meter
Canada-France-Hawaii Telescope (CFHT) atop Mauna Kea, Hawaii. MegaCam
uses forty 2048 x 4612 pixel CCDs, covering a $1^{o}$x $1^{o}$ field
of view with a resolution of 0.187 arcsec/pixel. The images used were
taken as part of the {}``Very Wide'' segment of CFHT Legacy Survey
(CFHTLS). Seven sets of observations were taken in MegaPrime $\gp$
filter ($\sim400-580$ nm) on either 2004 December 15-16 or 2005 January
16-17 , nineteen sets in the MegaPrime $\rp$ filter ($\sim550-700$
nm) were taken the nights of 2006 May 1- 2 and May 25-26.

Images from the VW segment of the survey were chosen for this study
because of its cadence: three images are taken of the same field at
approximately 45 minute intervals during the course of the first night,
followed by a single image of the same field the following night.
The large field of view of the camera means that 1) many asteroids
are seen on any given frame and 2) many of these can be followed up
successfully on the second night, which allows for somewhat improved
orbits, geocentric and heliocentric distances and hence sizes.

Data were obtained in both the CFHTLS $\gp$ and $\rp$ filters in
order to compare results at two wavelength ranges. The CFHTLS VW survey
also acquired images in the $\ip$ filter. However these were taken
far from opposition. As a result it proved much more difficult to
make accurate helio and geocentric distance determinations (even given
a detection on the second night), and we excluded them from our sample. 

The exposure times were 90 seconds for $\gp$ and 110 seconds in $\rp$.
Seeing sizes were 0.8'' and 1.1'' respectively for the 2 dates the
$\gp$ frames were taken. Limiting magnitude for 50\% probability
of three sigma detection of the $\gp$ frame with 1.1'' seeing is
23.0, the 90\% probability limit is 22.5. For the $\gp$ frames the
seeing was 1.0'' and 1.1'' for the two dates, and the limiting magnitude
for a three sigma detection was 21.75 and 22.25 for 90\% and 50 \%
respectively, for the night with the worse seeing. The limiting magnitude
was determined by calibrating on a set of images containing artificially
implanted sources moving at rates consistent with those of MBAs. The
images used for the seeding were real data images from the CFHTLS;
the artificial sources were implanted using the \noun{mkobjects}
function of IRAF \citep{tod86}. The information is used to set a
detection limit, which we choose to be 90\% completion (that is, 21.75
in $\rp$ and 22.5 in $\gp$). We base our further analysis only on
those objects brighter than the above limits.

The CFHTLS images were processed by the Elixir pipeline, which includes
bias and dark subtraction, flat-fielding and fringe subtraction. Photometric
corrections including colour terms are computed at this time. The
images are then processed by the Terapix data processing centre based
in Paris for fine astrometric correction to the USNO~B1.0 catalog
\citep{monlevcan03}. The cleaned images are then stored at the Canadian
Astronomical Data Centre, from which we retrieved them and began the
search for moving objects. 

The fields taken in the different filters were taken at different
times. No attempt was made to take images in both filters on the same
night, nor to follow particular asteroids for more than two nights.
As a result, the fields taken with different filters do not contain
the same asteroids (except possibly by chance). Thus the size distributions
determined in the $\gp$ and $\rp$ filters are for two statistically
similar samples of asteroids, rather than for the same sample as seen
through the two filters. Total survey areas were 7 fields ($\sim7$~square
degrees) in the $\gp$ and 19 fields ($\sim19$~square degrees) in
the $\rp$ band, with all fields taken within $\pm2$ degrees of the
ecliptic.

\subsection{Asteroid detection}

In order to detect MB asteroids in our images, Source Extractor \citep{berarn96}
was used to build a catalog of all sources more than 3 sigma above
the background, and provided the sources' positions (both x-y and
RA/Dec), magnitudes, full-width-half-max, as well as flags that described
sources that were saturated, truncated, blended with another, or located
on bad pixels. The earlier Terapix processing of the frames produces
photometric corrections for filter and airmass and these are applied
by Source Extractor in the calculation of the magnitudes.

Stationary objects are then removed; as are sources with the obvious
characteristics of cosmic rays. The remaining sources are then searched
for triplets moving within the appropriate range of angular rates.
Those detected are considered candidate one-night asteroid detections. 

The image areas surrounding each candidate are then blinked and a
human operator determines whether the candidate is real, or the result
of imperfect cosmic ray removal, variations in the image quality during
the night or other causes. Candidates not clearly visible and asteroidal
in appearance in all three frames are discarded. Those that remain
constitute our sample of one-night objects and will be subjected to
further analysis, both as to their size distribution and as to whether
or not they are seen on the second night's image. 

In order to determine whether or not the objects appear in the second
night's image, the motion of the candidate is extrapolated linearly
forward in time. If the position is determined to have moved out of
the field of view, the processing proceeds no further. If it is predicted
to fall within the second night's image, a blink frame of the section
of the second night's image around the predicted position is compared
to the same region taken the previous night. If blinking reveals an
object of appropriate magnitude near the appropriate location on the
second night, the object is deemed to have been detected on the second
night. A final consistency check is performed by computing an orbit
for the object based on the two nights of observations, and verifying
that the motion is reasonable and within the main belt (Trojan asteroids,
centaurs and Kuiper Belt objects are occasionally picked up). The
final catalog of one and two-night detections is what is analyzed
for its size-frequency distribution. 

We saw 686 main-belt objects only on a single night, and 839 on two
nights, or 272/414 and 245/594 one and two-nighters respectively in
the $\gp$/$\rp$ filters. We see a total of 1525 asteroids in both
filters, and 73 and 53 asteroids per square degree in $\gp$ and $\rp$
respectively.

\subsection{Orbital elements}

For the single-night detections, the arc was approximately 1.5 hours
long. In order to compute the semimajor axis and inclination, we used
Vaiasala's method based on the assumption that one observation was
taken at perihelion, taking the first and last observations, as described
by \cite{dub61}. A method proposed by Dubyago in that same work and
based on the assumption of a circular orbit was also tested. Comparisons
done using known asteroids with well-determined orbits revealed Vaiasala's
method to be somewhat superior for these objects with very short arcs.
For the two-night detections, Herget's method (as described in \cite{dan89})
was used, because of its slightly superior performance when tested
on observations of known asteroids. Herget's method requires estimates
of the geocentric distance of the body in question, however these
can be obtained fairly accurately given observational arcs of about
1 day for asteroids within the main-belt.

\subsection{Absolute magnitudes and diameters}

The absolute magnitude $H_{k}$ in filter $k$ is determined from
the apparent magnitude $m_{k}$ in appropriate filter from \begin{equation}
m_{k}=H_{k}+5\log_{10}(r\Delta)+P(\alpha)\end{equation}
 where $r$ and $\Delta$ are the heliocentric and geocentric distances
of the asteroid, $\alpha$ is the phase angle and $P(\alpha)$ is
the phase function. We use $P(\alpha)$ from \cite{bowhapdom89} with
a $G$ of 0.15. In all cases, the phase angle is small, ranging from
1.7 to 7.0 degrees with a mean of $3.9^{o}$. The rms errors in $r$
and $\Delta$(which are essentially equal and are strongly correlated
in our sample) were both 0.38 AU and 0.32 AU for the one and two night
detections respectively. We had hoped that the two night observations
would provide us with significantly improved $r$ and $\Delta$ accuracy
 however a longer arc, of order a week, is likely required to achieve
much improvement. Errors in $m_{k}$ were relatively small, and as
a result errors in $H_{k}$ (which ranged from 0.54 to 0.69 magnitudes
for the one and two night detections respectively) are dominated by
the uncertainties in $r$ and $\Delta$. 

The diameter estimate is derived from \cite{bowhapdom89}\begin{equation}
D=\frac{1347\times10^{-H_{k}/5}}{A_{k}^{1/2}}\end{equation}
 where $D$ is diameter in km and $A_{k}$ is albedo in filter $k$.
The uncertainly in $D$ receives nearly equal contributions from the
albedo and $H_{k}$ : the one sigma error is 0.36-0.4 km for the one
or two night detection.

The cumulative number distribution for main belt asteroids brighter
than an absolute magnitude $H_{k}$ (i.e. having a magnitude less
than $H_{k}$) can be approximated as \begin{equation}
\log N(<H_{k})=C+\gamma H_{k}.\end{equation}
 where $N$ is the cumulative number of asteroids, and $\gamma$ and
$C$ are constants, with $\gamma$ being the slope. Rewriting the
equation above in terms of diameter \begin{equation}
N(>D)\propto D^{-b}\end{equation}
 Here, the power-law index, $b$, corresponds to the slope of the
log $N$ vs. log $D$ plot, and is connected to the constant $\gamma$
by $b=5\gamma$. Using the method of \cite{yosnakwat03}, we will
use $b$ to express the slope of the cumulative size distribution
of asteroids. Note that the slope of the size-frequency distribution
is expressed in a variety of ways in the literature; a useful {}``translation
table'' can be found in Appendix A of \cite{obrgre05}.

\subsection{Previous work\label{sub:Previous-work}}

There are a few major surveys that have calculated cumulative size
distribution (CSD) slopes to which we can compare our own value. The
first is the Yerkes-McDonald Survey (YMS), which was the first (1951-1952)
systematic photographic survey with asteroid magnitudes based on a
photometric system. They found 1550 asteroids with a limiting magnitude
of 16.5. They calculated a CSD slope of $b=2.4$ for asteroids from
30-300 kilometers \citep{kuifuggeh58}. The next major survey, Palomar-Leiden,
was another photographic survey, performed in 1960, and which extended
the magnitude-frequency distribution to a magnitude of about 20. They
found over 2000 asteroids and calculated a slope of $b=1.8$ for asteroids
larger than 5 kilometers in diameter \citep{vanvanher70}. 

From 1992-1995 Spacewatch detected 59226 asteroids larger than 5 kilometers.
The limiting magnitude for this survey was about 21 in the visual
band, and yielded a CSD slope, again, of $b=1.8$ \citep{jedmet98}.
A survey of a relatively small field (15' square) by ISO at 12$\mu m$
saw 20 sources and deduced a shallow slope for smaller asteroids as
well, in this case $b=1.5$ \citep{teddes02}. A study of asteroid
sizes performed with archived frames from HST's WFPC2 camera taken
from 1994 to 1996 revealed 96 moving objects with apparent magnitudes
down to 24, or with diameters of 0.3 to 3 km \citep{evastapet98}.
This work found a slope of 1.2 to 1.3, even shallower than found by
previous investigators. 

The Sloan Digital Sky Survey (SDSS), carried out between 1998 and
2000, systematically mapped an enormous part of the sky and produced
detailed images allowing the determination of positions and absolute
magnitudes of many celestial bodies, including many asteroids. \citep{ivetabraf01}
used this survey to calculate a CSD slope for 13000 asteroids down
to a magnitude of 21.5 (in the R-band filter) and obtained a value
of $b=1.3$ for asteroids in the diameter range 0.4-5 kilometers.
The Sub-km Main-Belt Asteroid Survey (SMBAS) performed at the Subaru
telescope found 1111 asteroids down to a limiting magnitude of 24.4
and calculated the CSD, for asteroids between 0.5 and 1.0 kilometers,
to have $b=1.2$ \citep{yosnakwat03}. However, not all studies have
revealed a shallowing slope at smaller sizes. A recent report gives
a constant $b=1.9$ slope down to roughly 23 magnitude in V \citep{davglajed06}

However, it does appear that sub-km asteroids display a somewhat shallower
CSD slope than the largest ones: this implies a deficit in the smaller
asteroids, indicative of some changing physics as we move into the
regime where collisional fragmentation becomes more dependent on internal
strength and less on gravity.

\section{Results}

The cumulative distributions of asteroid diameters in our sample are
shown in Fig 1, with the assumption that all asteroids have an albedo
of 0.09. The shaded region below each observed distribution indicates
the difference between the observed CSD (the heavy line) and that
which only includes asteroids brighter than our 90\% completion limit.
Thus the thickness of the shaded region gives us a measure of how
much our sample is affected by incompleteness. In fitting slopes to
the observed distribution, we only use those points where the shaded
area is less than 10\% of the height of the observed distribution.
Put another way, we only fit those points where objects fainter than
our completeness limit contribute less than 10\% to the height of
the distribution at a given point, to eliminate a skewing of the distribution
due to incompleteness.

\plotone{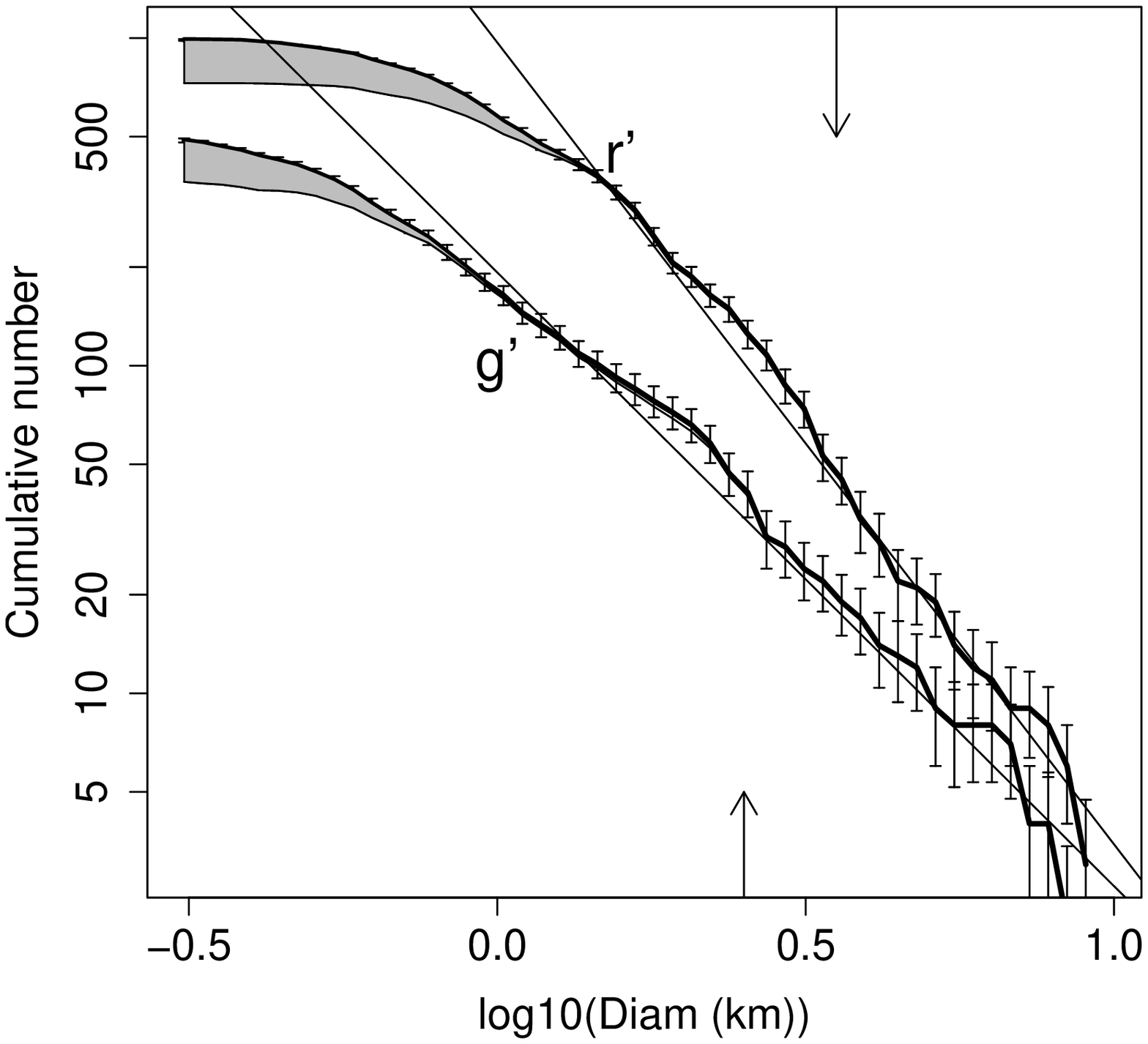}

The error bars on the CSDs are determined by a standard bootstrap
process \citep{efr82}. Using our sample of diameter measurements,
each with an individually computed uncertainty, we generated one hundred
statistically similar distributions by a Monte Carlo process under
the assumption that the errors are distributed in a Gaussian fashion.
The plotted error bars in the figures represent one standard deviation
as computed by the bootstrap process at each point.

The least-squares lines shown in the figures are fitted only to data
points where the observed CSD and that based on the 90\% completion
limit differ by less than 10\%. The data points are weighted by 1/sigma
during this fit to properly account for the larger error bars in the
larger diameter region of the plot, though an unweighted fit produces
similar results. We note however the distributions do not seem particularly
well fit by a straight line in this range; there are features which
deviate from the line by more than the error bars in Figure 1. Deviations
from a pure-power law slope for asteroids CSDs are now well-known
and have been discussed by many authors, for example \citet{celzapfar91,durder97,durgrejed98,obrgre03}. 

The difference between the slopes in the two filters is quite clear
in Figure 1. The best-fit slope for the $\gp$ sample is $b=1.87\pm0.05$
while that for the $\rp$ filter is $b=2.45\pm0.07$. We also note
that the slope difference is not simply due to our relatively fine
binning. A coarser binning, shown in Figure 2, produces least-squares
fit slopes which show the same trend ($1.94\pm0.15$ and $2.23\pm0.11$).
The difference in the slopes is not quite as distinct and has larger
uncertainties as we are fitting the line to relatively few points
(we continue to exclude those beyond our 90\% completeness limits),
yet the slopes still differ by about two sigma.

\plotone{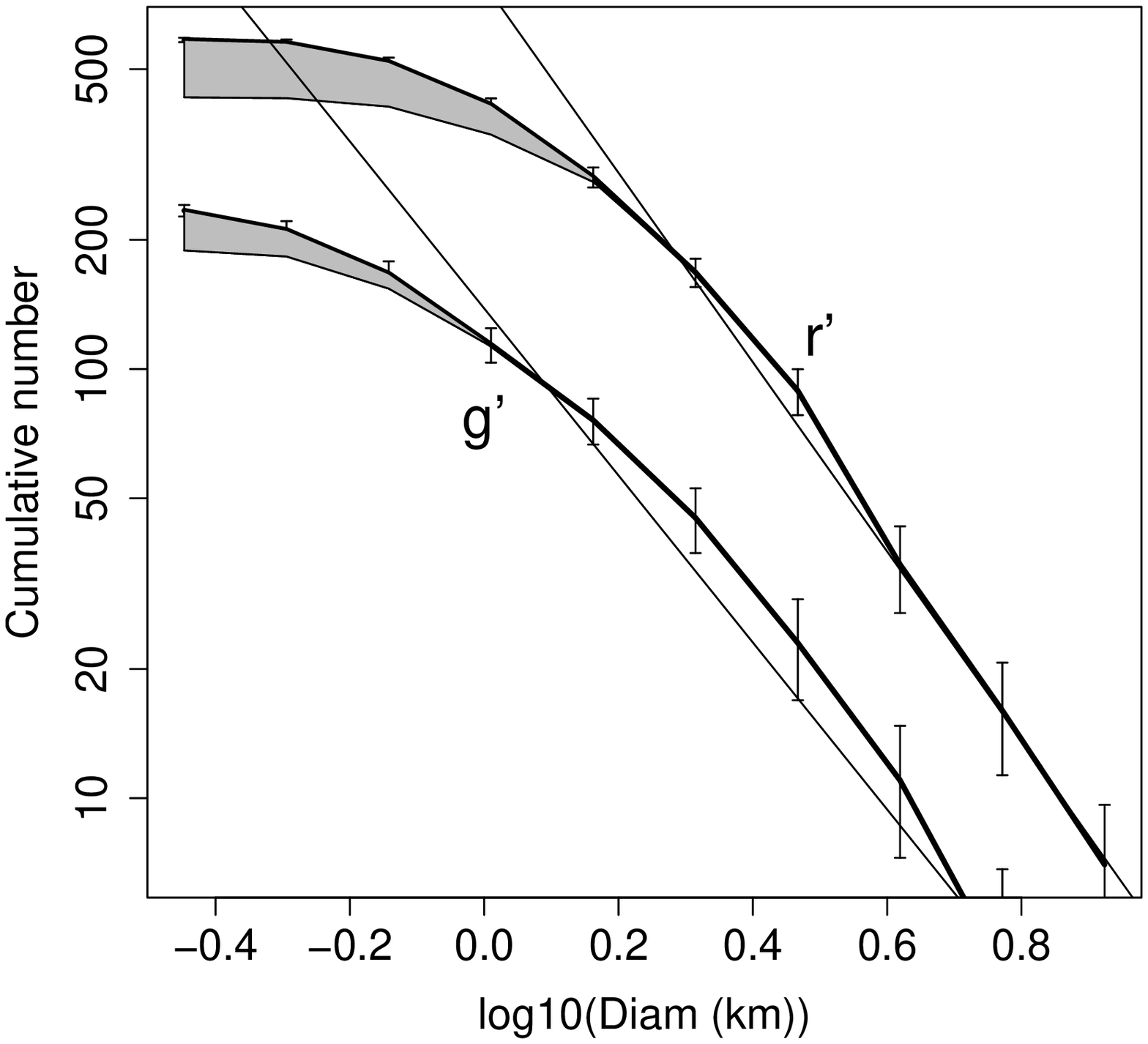}

Before discussing the difference between the $\gp$ and $\rp$ slopes,
we first note that both show evidence for at least one change in the
power law slope, at roughly 2.5 km in $\gp$ and 3.5 km in $\rp$
(see Figure 1). The location of this change in the slope {}``knee''
as seen in our sample corresponds roughly to that seen in other determinations
of the CSD, for example, that in Figure 1 of \citet{obrgre05}. The
causes of such deviations of the asteroid size distribution from a
smooth power-law remain under study, but almost certainly reflect
the influence of size-dependent processes in the asteroid belt.

A slope calculated only from asteroids smaller than this {}``knee''
yields a slope of $b=1.35\pm0.02$ and $1.79\pm0.07$ in the visible
and red respectively, shallower than the overall slope in each case.
Owing to the small number of asteroids larger than the knee in our
sample we did not calculate the slope for these bodies on their own.
Their effect on the slope when excluded from the least-squares fit
is sufficient to show their tendency towards a steeper slope than
the smaller asteroids. Thus both filters show a reduced slope for
smaller bodies, which is consistent with other observational results
where, despite variations, a general trend towards shallower slopes
at smaller diameters is evident. This trend has been associated with
size-dependent depletion of small asteroids because 1) they may acquire
higher velocities during collisions, 2) they are subject to larger
Yarkovsky drifts and 3) they have lower strengths per unit mass.

Notably, the slopes presented in the previous paragraph span the range
of values quoted for asteroids of various sizes (see section \ref{sub:Previous-work}),
illustrating the danger of fitting a single line to a distribution
whose character is more complicated. Our interpretation of why our
results show slopes generally steeper than those reported earlier
at these sizes is simply that a pure power law is not a very good
fit to the size distribution of main belt asteroids. We have examined
only about one and a half orders of magnitude in diameter, a relatively
narrow range and roughly the size of the {}``waves'' expected in
the distribution due to size-dependent processes (See Figure 6 in
\citet{durder97}). Though an overarching power-law component is clearly
present in the size distribution of MB asteroids, deviations from
a pure power law, which have both been seen observationally by many
authors and which are expected theoretically, clearly make a simple
one-parameter characterization of the entire size distribution unworkable.

Of more interest are the differing CSD slopes for asteroids viewed
in different filters, though such a difference is perhaps not unexpected.
As one moves towards smaller sizes, the strength becomes composition-dependent
as gravity becomes less of a factor, and it is known that there are
widely differing compositions across the asteroid belt. Simply speaking,
each sample should contain different proportions of asteroids with
different colours, and hence compositions and internal properties.
Since internal properties being to dominate the bodies' strength,
the collisionally-induced size distribution might be expected to be
different.

We are not aware of a colour or filter dependence in the size distribution
having been reported before. This might be explained by the relatively
few earlier studies that could reach this size range, or in the cases
of those that did, by an absence of colour information. We note that
the result of steeper slopes for $\rp$ versus $\gp$ that we find
runs counter to earlier work, where studies performed in the red typically
show shallower slopes than those performed in the visible. However,
there is also a correlation between the time at which the studies
were performed and the slopes observed. Earlier studies were done
in the visible and saw larger bodies than later studies typically
done in the red, making it difficult to distinguish the effects of
colour at different sizes from these previous results.

In order to examine the difference in slope with filter more extensively,
we split our sample into three semi-major axis regions, following
\citet{yosnak04}. The three zones used are the inner $(2.0<a({\rm AU)\le2.6)}$,
middle $(2.6<a({\rm AU)\le3.0)}$, and outer $(3.0<a({\rm AU)\le3.5)}$
zones. Our semimajor axis determination has an uncertainty of 0.3
AU (based on the comparison of our calculations with known asteroids
(observed by chance), so this division is a rough one, but helps reveal
whether the different slopes remain evident in subsamples of our data
set. The diameter distributions for each region are shown in Figures
3, 4 and 5.

\plotone{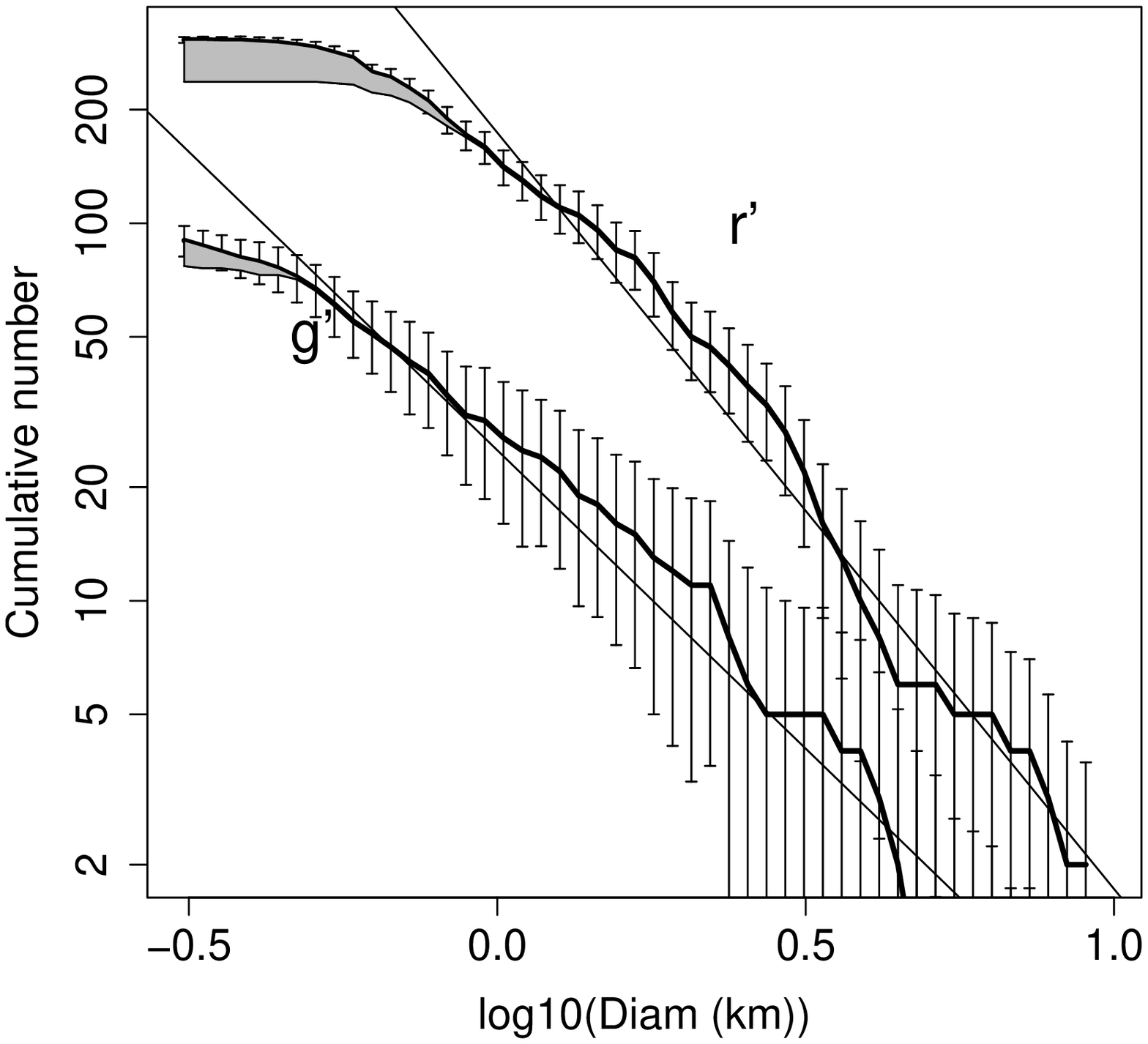}
\plotone{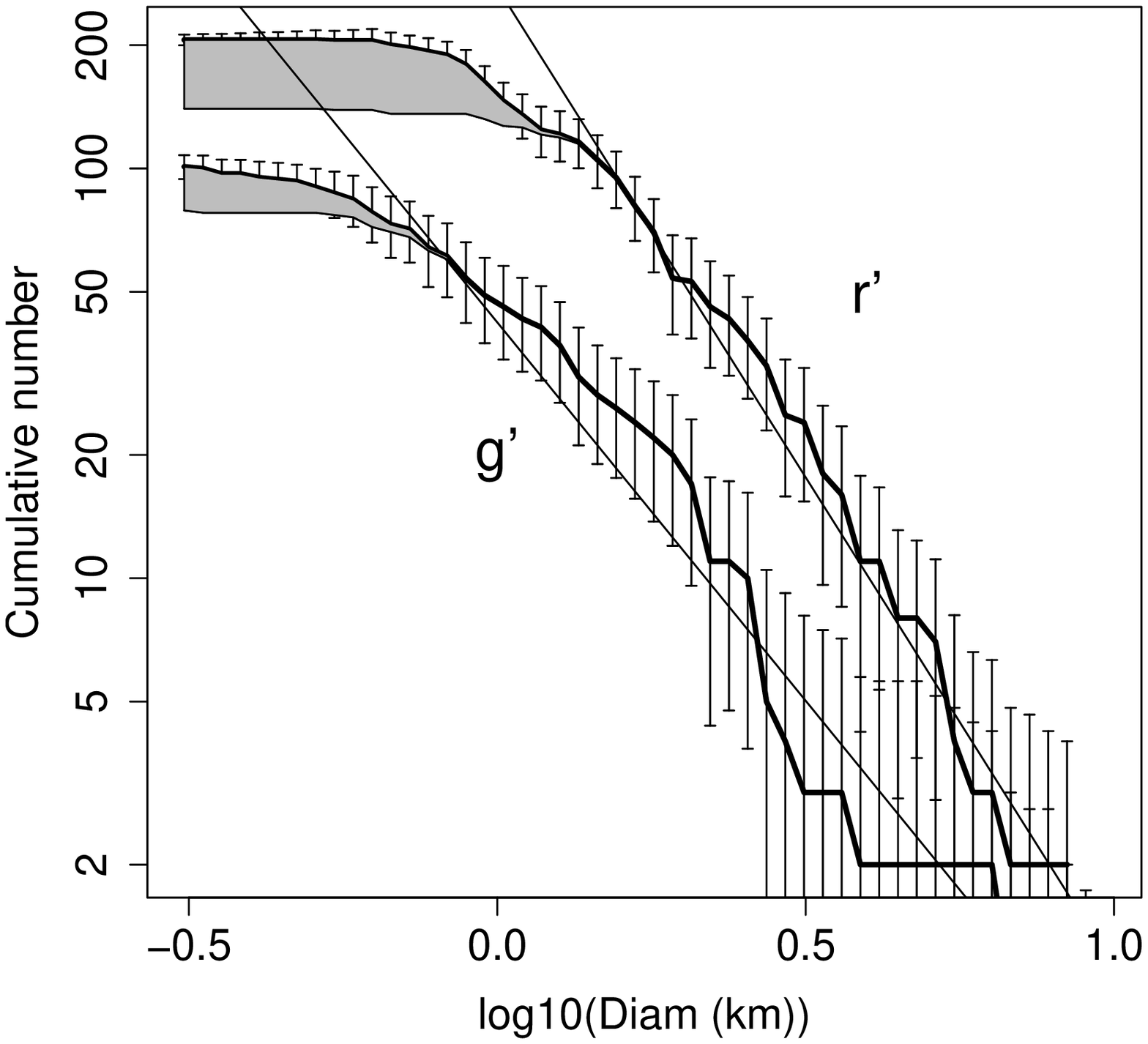}
\plotone{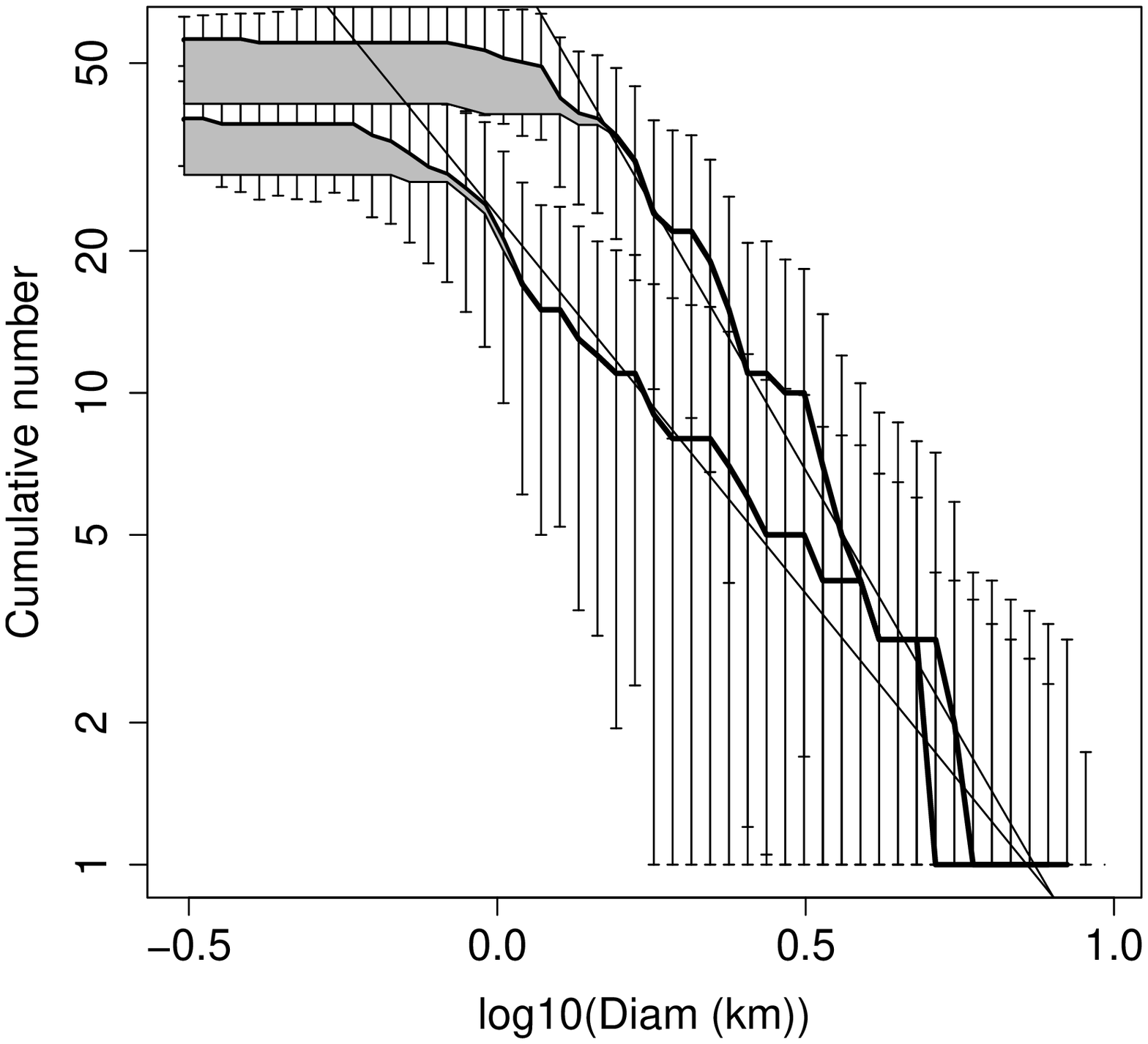}

In these subsamples, slope difference persists, though the slopes
are not constant across the MB. The slope of the $\rp$ distribution
shows strong variations with semimajor axis (see Table 1). The least-squares
fit for the middle region (Fig. 4) has the steepest slope ($b=2.39\pm0.07)$
with the outer region (Fig. 5) next ($b=2.25\pm0.08$). The inner
region (Fig 3) has the shallowest slope in $\rp$ ($b=2.00\pm0.05)$.
The $\gp$ distributions has a less dramatic but similar trend, also
showing the steepest slope in the middle belt $(b=1.85\pm0.06)$ while
the inner and outer belt are similar at $b=1.58\pm0.06$ and $b=1.60\pm0.07$
respectively. Differences across the asteroid belt are not unexpected
owing to the well-known compositional variations with semi-major axis
\citep{grated82,motcarlaz03}, but we note that the shallower slope
for asteroids seen in the $\gp$ versus the $\rp$ filter is present
in each our subsamples of the main belt.

\begin{center}\begin{tabular}{|c|c|c|c|c|}
\hline 
Filter&
range&
$b$&
sigma&
$N$\tabularnewline
\hline
\hline 
$\gp$&
all&
1.87&
0.05&
185\tabularnewline
\hline 
$\rp$&
all&
2.44&
0.07&
423\tabularnewline
\hline 
$\gp$&
inner&
1.58&
0.06&
77\tabularnewline
\hline 
$\rp$&
inner&
2.00&
0.05&
238\tabularnewline
\hline 
$\gp$&
middle&
1.85&
0.06&
79\tabularnewline
\hline 
$\rp$&
middle&
2.39&
0.07&
143\tabularnewline
\hline 
$\gp$&
outer&
1.60&
0.07&
29\tabularnewline
\hline 
$\rp$&
outer&
2.25&
0.08&
42\tabularnewline
\hline
\end{tabular}\end{center}

In order to examine these findings in more detail, we also compute
the slope for those asteroids with sizes smaller than the {}``knee''
in the distributions mentioned earlier. This allows us to work in
a region where the error bars are smaller, and puts us more firmly
in the regime where strength depends more on internal composition,
and thus where colour-related effects may be stronger. Though we would
expect the absolute slopes of these smaller-diameter sections of the
distributions to be shallower (as discussed earlier in this section),
they can be examined to see whether they show the same trend. 

These more finely divided subsamples show the same qualitative behaviour
seen earlier. Both 1) the significantly higher slope in the red versus
the visible across the belt, and 2) in $\rp$, higher slopes in the
middle/outer belt, are seen. The results are summarized in Table 2.
An overall shallower slope at these sizes, expected from our examination
of the complete samples is seen, but the filter-related differences
persist. There are other small differences. The main difference is
that the slope of these subsamples in $\rp$ in the middle and outer
belt are equal ($b=1.8\pm0.07$ and $1.81\pm0.09$, Table 2), whereas
they differ in the complete sample ($b=2.39\pm0.07$ and $2.25\pm0.08$,
Table 1). However, the outer region is where we have the fewest objects
and hence is likely to be the least reliable in terms of slope determination.

\begin{center}\begin{tabular}{|c|c|c|c|c|}
\hline 
Filter&
a range&
b&
sigma&
N\tabularnewline
\hline
\hline 
$\gp$&
all&
1.35&
0.02&
167\tabularnewline
\hline 
$\rp$&
all&
1.91&
0.08&
384\tabularnewline
\hline 
$\gp$&
inner&
1.20&
0.03&
71\tabularnewline
\hline 
$\rp$&
inner&
1.58&
0.06&
224\tabularnewline
\hline 
$\gp$&
middle&
1.39&
0.07&
73\tabularnewline
\hline
$\rp$&
middle&
1.80&
0.07&
124\tabularnewline
\hline
$\gp$&
outer&
1.31&
0.05&
23\tabularnewline
\hline
$\rp$&
outer&
1.81&
0.09&
36\tabularnewline
\hline
\end{tabular}\end{center}

Thus we conclude that there is a real difference in the slopes of
the CSDs as seen in the two filters, and that this difference appears
strongest in the middle and outer asteroid belt, but somewhat less
pronounced in the inner belt. The slope of the CSD in the $\gp$ filter
shows weak variations throughout the belt, while the $\rp$ distribution
shows larger changes, particularly from the inner to the middle/outer
belt. 

Unfortunately there we do not have enough information to distinguish
between the effects of colour, albedo, size, age and strength in this
system, making a determination of the cause of the different slopes
a difficult task. However, the persistence of the slope differences
when the sample is subdivided gives us some confidence that the result
is real. At the very least, it seems likely that the asteroid size
distribution is colour-dependent within certain regions of the main
belt. 

A number of different scenarios could be imagined as causes for the
slope differences, most tied to the known composition gradient across
the main belt \citep{grated82,motcarlaz03}. The difference in the
$\gp$ and $\rp$ slopes in middle and outer belt might be interpreted
as evidence for two differently coloured asteroid complexes, with
the red ($\rp$) sample containing more small asteroids for each larger
one indicating perhaps that the red asteroids are weaker per unit
mass. The weaker slope difference in the inner belt may mean that
there is only one dominant asteroid complex here, and we are seeing
it in both filters. 

Despite the temptation to link the samples as seen through the different
filters with particular asteroid types, it is clear that we do not
have enough information to make unique associations. We do not have
true colour information on any asteroids observed, as none of the
the bodies were seen through both filters. Other considerations include
albedo, which also plays a role in selecting our samples, and which
we have not considered here. One could imagine an age-dependent component
as well, as asteroid weathering produces redder colours but is not
expected to affect C and S type asteroids equally. More observations,
with more spectral information information is required to determine
the cause of the colour dependence on size across the main asteroid
belt.

\section{Conclusions}

We detected 517 and 1008 main belt asteroids in the $\gp$ and $\rp$
filter respectively in CFHTLS MegaPrime/MegaCam images using Source
Extractor. We used the Vaiasala and Herget techniques to calculate
the orbital elements from one or two nights' observations respectively.
We then used the average apparent magnitude of our asteroids and converted
first to absolute magnitude using an assumed albedo of 0.09, then
to diameter. We used the diameters to create a cumulative size distribution
(CSD) plot of our asteroids and determine CSD slopes for various subsets
of our data. 

We found a general trend towards shallower slopes in the CSD as we
moved towards smaller diameters, as has typically been found by other
researchers. Our overall best fit slopes are typically higher than
reported previously, which we attribute to the sensitivity of the
slope determination to deviations from a pure power law, and the narrow
range of diameters in our sample. We determine that the overall size
distribution does show a filter dependence over the size range examined,
indicating that smaller asteroids in the sample seen in the $\rp$
filter are relatively more abundant than those we detect in $\gp.$
This difference is weaker in the inner belt, but prominent in the
middle and outer parts of the belt. We conclude that there is evidence
for a colour dependence in the size-distribution of asteroids in the
0.3 - 10 km diameter range, a variation whose strength differs across
of the belt, though further investigation is required to determine
the underlying cause of the observed difference.

\acknowledgements{

This research was performed in part with support from the National
Science and Engineering Research Council of Canada. This work is based
on observations obtained with MegaPrime/MegaCam, a joint project of
CFHT and CEA/DAPNIA, at the Canada-France-Hawaii Telescope (CFHT)
which is operated by the National Research Council (NRC) of Canada,
the Institut National des Sciences de l'Univers of the Centre National
de la Recherche Scientifique (CNRS) of France, and the University
of Hawaii. This work is based in part on data products produced at
TERAPIX and the Canadian Astronomy Data Centre as part of the Canada-France-Hawaii
Telescope Legacy Survey, a collaborative project of NRC and CNRS.
This research used the facilities of the Canadian Astronomy Data Centre
operated by the National Research Council of Canada with the support
of the Canadian Space Agency.
}

\bibliographystyle{natbib}
\bibliography{Wiegert}

\begin{thebibliography}{{Yoshida} and {Nakamura}(2004)}

\bibitem[{Benz} and {Asphaug}(1999)]{benasp99}
{Benz}, W. and {Asphaug}, E.
\newblock November 1999.
\newblock {Catastrophic Disruptions Revisited}.
\newblock {\em Icarus} 142:5--20.

\bibitem[{Bertin} and {Arnouts}(1996)]{berarn96}
{Bertin}, E. and {Arnouts}, S.
\newblock 1996.
\newblock {SExtractor: Software for source extraction.}
\newblock {\em Astron. Astrophys. Suppl.} 117:393--404.

\bibitem[Bowell {\it et~al.}(1989)]{bowhapdom89}
Bowell, E., Hapke, B., Domingue, D., Lumme, K., Peltoniemi, J.,  and Harris, A.
\newblock 1989.
\newblock Application of photometric model to asteroids.
\newblock In {\em Asteroids II}, R.~Binzel, T.~Gehrels, and M.~Matthews (eds.),
  pages 524--556. Tucson: University of Arizona Press.

\bibitem[Cellino {\it et~al.}(1991)]{celzapfar91}
Cellino, A., Zappala, V.,  and Farinella, P.
\newblock 1991.
\newblock The size distribution of main-belt asteroids from {IRAS}.
\newblock {\em MNRAS} 253:561--574.

\bibitem[Danby(1989)]{dan89}
Danby, J. M.~A.
\newblock 1989.
\newblock {\em Fundamentals of Celestial Mechanics}.
\newblock Richmond VA: Willmann-Bell.

\bibitem[{Davis} {\it et~al.}(1994)]{davryafar94}
{Davis}, D.~R., {Ryan}, E.~V.,  and {Farinella}, P.
\newblock August 1994.
\newblock {Asteroid collisional evolution: Results from current scaling
  algorithms}.
\newblock {\em Plan. Space Sci.} 42:599--610.

\bibitem[{Davis} {\it et~al.}(2006)]{davglajed06}
{Davis}, D.~R., {Gladman}, B., {Jedicke}, R.,  and {Williams}, G.
\newblock 2006.
\newblock {The Sub-Kilometer Asteroid Diameter Survey}.
\newblock In {\em AAS/Division for Planetary Sciences Meeting Abstracts}, page
  53.01.

\bibitem[{Dohnanyi}(1969)]{doh69}
{Dohnanyi}, J.~S.
\newblock 1969.
\newblock {Collisional model of asteroids and their debris}.
\newblock {\em J. Geophys. Res.} 74:2531--2554.

\bibitem[Dubyago(1961)]{dub61}
Dubyago, A.~D.
\newblock 1961.
\newblock {\em The Determination of Orbits}.
\newblock New York: MacMillian.

\bibitem[{Durda} and {Dermott}(1997)]{durder97}
{Durda}, D.~D. and {Dermott}, S.~F.
\newblock November 1997.
\newblock {The Collisional Evolution of the Asteroid Belt and Its Contribution
  to the Zodiacal Cloud}.
\newblock {\em Icarus} 130:140--164.

\bibitem[{Durda} {\it et~al.}(1998)]{durgrejed98}
{Durda}, D.~D., {Greenberg}, R.,  and {Jedicke}, R.
\newblock October 1998.
\newblock {Collisional Models and Scaling Laws: A New Interpretation of the
  Shape of the Main-Belt Asteroid Size Distribution}.
\newblock {\em Icarus} 135:431--440.

\bibitem[Efron(1982)]{efr82}
Efron, B.
\newblock 1982.
\newblock {\em The Jackknife, the Bootstrap and Other Resampling Plans}.
\newblock Philadelphia: Society for Industrial and Applied Mathematics.

\bibitem[{Evans} {\it et~al.}(1998)]{evastapet98}
{Evans}, R.~W., {Stapelfeldt}, K.~R., {Peters}, D.~P., {Trauger}, J.~T.,
  {Padgett}, D.~L., {Ballester}, G.~E., {Burrows}, C.~J., {Clarke}, J.~T.,
  {Crisp}, D., {Gallagher}, J.~S., {Griffiths}, R.~E., {Grillmair}, C.,
  {Hester}, J.~J., {Hoessel}, J.~G., {Holtzmann}, J., {Krist}, J., {McMaster},
  M., {Meadows}, V., {Mould}, J.~R., {Ostrander}, E., {Sahai}, R., {Scowen},
  P.~A., {Watson}, A.~M.,  and {Westphal}, J.
\newblock 1998.
\newblock {Asteroid Trails in Hubble Space Telescope WFPC2 Images: First
  Results}.
\newblock {\em Icarus} 131:261--282.

\bibitem[{Farinella} {\it et~al.}(1982)]{farpaozap82}
{Farinella}, P., {Paolicchi}, P.,  and {Zappala}, V.
\newblock December 1982.
\newblock {The asteroids as outcomes of catastrophic collisions}.
\newblock {\em Icarus} 52:409--433.

\bibitem[{Gradie} and {Tedesco}(1982)]{grated82}
{Gradie}, J. and {Tedesco}, E.
\newblock 1982.
\newblock {Compositional structure of the asteroid belt}.
\newblock {\em Science} 216:1405--1407.

\bibitem[{Housen} and {Holsapple}(1990)]{houhol90}
{Housen}, K.~R. and {Holsapple}, K.~A.
\newblock March 1990.
\newblock {On the fragmentation of asteroids and planetary satellites}.
\newblock {\em Icarus} 84:226--253.

\bibitem[{Housen} and {Holsapple}(1999)]{houhol99}
{Housen}, K.~R. and {Holsapple}, K.~A.
\newblock November 1999.
\newblock {Scale Effects in Strength-Dominated Collisions of Rocky Asteroids}.
\newblock {\em Icarus} 142:21--33.

\bibitem[{Ivezic} {\it et~al.}(2001)]{ivetabraf01}
{Ivezic}, Z., {Tabachnik}, S., {Rafikov}, R., {Lupton}, R.~H., {Quinn}, T.,
  {Hammergren}, M., {Eyer}, L., {Chu}, J., {Armstrong}, J.~C., {Fan}, X.,
  {Finlator}, K., {Geballe}, T.~R., {Gunn}, J.~E., {Hennessy}, G.~S., {Knapp},
  G.~R., {Leggett}, S.~K., {Munn}, J.~A., {Pier}, J.~R., {Rockosi}, C.~M.,
  {Schneider}, D.~P., {Strauss}, M.~A., {Yanny}, B., {Brinkmann}, J., {Csabai},
  I., {Hindsley}, R.~B., {Kent}, S., {Lamb}, D.~Q., {Margon}, B., {McKay},
  T.~A., {Smith}, J.~A., {Waddel}, P., {York}, D.~G.,  and {the SDSS
  Collaboration}.
\newblock 2001.
\newblock {Solar System Objects Observed in the Sloan Digital Sky Survey
  Commissioning Data}.
\newblock {\em AJ} 122:2749--2784.

\bibitem[{Jedicke} and {Metcalfe}(1998)]{jedmet98}
{Jedicke}, R. and {Metcalfe}, T.~S.
\newblock 1998.
\newblock {The Orbital and Absolute Magnitude Distributions of Main Belt
  Asteroids}.
\newblock {\em Icarus} 131:245--260.

\bibitem[{Kuiper} {\it et~al.}(1958)]{kuifuggeh58}
{Kuiper}, G.~P., {Fugita}, Y.~F., {Gehrels}, T., {Groeneveld}, I., {Kent}, J.,
  {van Biesbroeck}, G.,  and {van Houten}, C.~J.
\newblock 1958.
\newblock {Survey of Asteroids.}
\newblock {\em Astrophys. J. Suppl.} 3:289--427.

\bibitem[{Monet} {\it et~al.}(2003)]{monlevcan03}
{Monet}, D.~G., {Levine}, S.~E., {Canzian}, B., {Ables}, H.~D., {Bird}, A.~R.,
  {Dahn}, C.~C., {Guetter}, H.~H., {Harris}, H.~C., {Henden}, A.~A., {Leggett},
  S.~K., {Levison}, H.~F., {Luginbuhl}, C.~B., {Martini}, J., {Monet},
  A.~K.~B., {Munn}, J.~A., {Pier}, J.~R., {Rhodes}, A.~R., {Riepe}, B., {Sell},
  S., {Stone}, R.~C., {Vrba}, F.~J., {Walker}, R.~L., {Westerhout}, G.,
  {Brucato}, R.~J., {Reid}, I.~N., {Schoening}, W., {Hartley}, M., {Read},
  M.~A.,  and {Tritton}, S.~B.
\newblock February 2003.
\newblock {The USNO-B Catalog}.
\newblock {\em AJ} 125:984--993.

\bibitem[{Moth{\'e}-Diniz} {\it et~al.}(2003)]{motcarlaz03}
{Moth{\'e}-Diniz}, T., {Carvano}, J.~M.~{\'A}.,  and {Lazzaro}, D.
\newblock March 2003.
\newblock {Distribution of taxonomic classes in the main belt of asteroids}.
\newblock {\em Icarus} 162:10--21.

\bibitem[{O'Brien} and {Greenberg}(2003)]{obrgre03}
{O'Brien}, D.~P. and {Greenberg}, R.
\newblock August 2003.
\newblock {Steady-state size distributions for collisional populations:
  analytical solution with size-dependent strength}.
\newblock {\em Icarus} 164:334--345.

\bibitem[{O'Brien} and {Greenberg}(2005)]{obrgre05}
{O'Brien}, D.~P. and {Greenberg}, R.
\newblock November 2005.
\newblock {The collisional and dynamical evolution of the main-belt and NEA
  size distributions}.
\newblock {\em Icarus} 178:179--212.

\bibitem[{Tedesco} and {Desert}(2002)]{teddes02}
{Tedesco}, E.~F. and {Desert}, F.-X.
\newblock 2002.
\newblock {The Infrared Space Observatory Deep Asteroid Search}.
\newblock {\em AJ} 123:2070--2082.

\bibitem[{Tody}(1986)]{tod86}
{Tody}, D.
\newblock 1986.
\newblock {The IRAF Data Reduction and Analysis System}.
\newblock In {\em Proc. SPIE Instrumentation in Astronomy VI}, D.~L. {Crawford}
  (ed.), page 733.

\bibitem[{van Houten} {\it et~al.}(1970)]{vanvanher70}
{van Houten}, C.~J., {van Houten-Groeneveld}, I., {Herget}, P.,  and {Gehrels},
  T.
\newblock 1970.
\newblock {The Palomar-Leiden survey of faint minor planets}.
\newblock {\em Astron. Astrophys. Suppl.} 2:339--448.

\bibitem[{Yoshida} and {Nakamura}(2004)]{yosnak04}
{Yoshida}, F. and {Nakamura}, T.
\newblock 2004.
\newblock {Basic nature of sub-km main-belt asteroids: their size and spatial
  distributions}.
\newblock {\em Advances in Space Research} 33:1543--1547.

\bibitem[{Yoshida} {\it et~al.}(2003)]{yosnakwat03}
{Yoshida}, F., {Nakamura}, T., {Watanabe}, J.-I., {Kinoshita}, D., {Yamamoto},
  N.,  and {Fuse}, T.
\newblock 2003.
\newblock {Size and Spatial Distributions of Sub-km Main-Belt Asteroids}.
\newblock {\em Publ. Astr. Soc. Japan} 55:701--715.

\end{thebibliography}

\section{Figure and table captions}
Figure 1. The heavy lines are the cumulative size distributions of
main-belt asteroids as detected in the $\rp$ and $\gp$ filters.
The shaded area indicates the difference between the observed distribution
and that in which we exclude all objects whose apparent magnitude
is below our 90\% completeness limit. The straight lines are the weighted
least-squares fit slopes to the size distribution, including only
those points where our completeness is above 90\%. The locations of
the slope changes at diameters of $\sim$2.5 ($\gp$) and $\sim$3.5
($\rp$) km are indicated by the arrows (see text).

Figure 2. The CSD with larger bin sizes.

Figure 3: Diameter distribution for the inner section of the belt
(2 < a <2.6 AU)

Figure 4: Diameter distribution for the middle section of the belt
(2.6 < a < 3.0 AU)

Figure 5: Diameter distribution for the outer section of the belt
(3.0 < a < 3.5 AU)

Table 1: Slopes across all sizes in different regions of the asteroid
belt. $N$ is the number of objects in the sample.

Table 2: Slopes for sizes smallest sizes ($D<2.5-3.5$ km, see the
text for more details).

\end{document}